# Excitation energy dependence of Raman spectra of few-layer WS$_2$

*Jinho Yang, Jae-Ung Lee and Hyeonsik Cheong*

Department of Physics, Sogang University, Seoul 04107, Korea

**Corresponding author:** E-mail: hcheong@sogang.ac.kr.

**ABSTRACT:** Raman spectra of few-layer WS$_2$ have been measured with up to seven excitation energies, and peculiar resonance effects are observed. The two-phonon acoustic phonon scattering signal close to the main $E_{2g}^1$ peak is stronger than the main peaks for excitations near the A or B exciton states. The low-frequency Raman spectra show a series of shear and layer-breathing modes that are useful for determining the number of layers. In addition, *hitherto* unidentified peaks (X$_1$ and X$_2$), which do not seem to depend on the layer thickness, are observed near resonances with exciton states. The polarization dependences of the two peaks are different: X$_1$ vanishes in cross polarization, but X$_2$ does not. At the resonance with the A exciton state, the Raman-forbidden, lowest-frequency shear mode for odd number of layers appears as strong as that for the allowed case of even number of layers. This mode also exhibits a strong Breit-Wigner-Fano line shape and an anomalous polarization behavior at this resonance.

**KEYWORDS**

WS$_2$, tungsten disulfide, Raman spectroscopy, exciton resonance, interlayer vibration mode

# 1. Introduction

Two-dimensional semiconductors, including transition metal dichalcogenides (TMDs) such as $MoS_2$, $MoSe_2$, and $WS_2$ are attracting much interest owing to unique 2-dimensional physical phenomena and superior opto-electronic properties that are advantageous for future devices. Valley polarization [1–4] and strong electron-hole interaction due to reduced dielectric screening [5–7] lead to several novel phenomena such as the valley Hall effect and observation of exciton-polaritons [8–10]. High electrical mobility and a large on/off ratio as well as large oscillator strength for optical transitions make them suitable for transistors [11,12], photodetectors [13,14], solar cells [15], biosensors [16] and light-emitting-devices [17,18] compatible with flexible electronics. Because of the strong electron-hole interaction, excitonic features play important roles in determining the optical properties of these materials. For example, in the absorption spectrum, the excitonic features are most prominent: the A exciton state associated with the bottom of the conduction band and the top of the valence band at the K and K′ points of the Brillouin zone and the B exciton state that is associated with the spin-orbit-split valence band are commonly observed [5,19,20]. In addition, the C exciton state associated with a higher energy band gap is observed [21–23]. The energetic separation between the A and B exciton states is determined by the spin-orbit interaction and is about 0.15 ~ 0.45 eV.

Raman spectroscopy is widely used to determine the number of layers and other important properties of layered 2-dimensional materials [24–29]. For TMD's, care should be taken because many of the commonly used excitation lasers are in resonance with one of the exciton states. For example, the 632.8 nm line of a He-Ne laser is closer to the resonance with the A and B excitons of $MoS_2$ [20,21,30], and many anomalous behaviors are observed when this laser line is used as the excitation source [28,31,32]. The resonance Raman effects, therefore, have become an important issue, and the correlation of anomalous resonance effects with different excitonic states is very interesting [23,33–35]. However, the contributions of A and B exciton states tend to be mixed in $MoS_2$ and $MoSe_2$ since the spin-orbit splitting is not large enough. $WS_2$ is a particularly useful because of a large spin-orbit splitting (~0.4 eV), which makes it relatively easy to probe the exciton-related phenomena for A and B excitons separately [5]. The conduction band also shows a small spin-orbit splitting. In the case of $MoS_2$, the higher valence band and the lower conduction band have the same spin, and so the exciton associated with the lowest energy transition is optically active (bright exciton). However, in the case of $WS_2$, because the spin-orbit splitting in the conduction band has an opposite sign, the lowest-energy exciton is not optically active (dark exciton) [36]. Therefore, the presence of the dark exciton state which has a slightly different energy than the bright exciton state should be taken into account in interpreting optical effects in this energy range. In this work, we measured the Raman spectra of mono- and few-layer $WS_2$ using up to seven excitation energies, some of which are in resonance with the A, B, or C exciton. Resonance effects are most pronounced in the low-frequency region in which the rigid shear and layer-breathing modes appear. Several new features are found and identified in the low-frequency Raman spectra of few-layer $WS_2$.

## 2. Methods

Few-layer $WS_2$ samples were prepared by mechanical exfoliation from $WS_2$ flakes (HQ Graphene) on Si substrates with a 285 nm $SiO_2$ layer. We identified the number of layers by comparing the optical contrast, PL, and Raman measurements [29,37–39]. The Raman measurements were performed in ambient conditions with seven excitation sources: the 325 and 441.6 nm (3.82 and 2.81 eV) lines of a He-Cd laser; the 457.9, 488, 514.5 nm (2.71, 2.54 and 2.41 eV) lines of an Ar ion laser; the 532 nm (2.33 eV) line of a diode-pumped-solid-state laser; and the 632.8 nm (1.96 eV) line of a He-Ne laser. A 50× objective lens (0.8 N.A.) was used to focus the laser beam onto the sample. The scattered light was collected by the same objective lens (backscattering geometry) and dispersed with a Horiba iHR550 spectrometer (2400 grooves/mm). We used reflective volume holographic filters (Ondax and OptiGrate) to prevent the laser line from entering the spectrometer. To avoid local heating, the laser power was kept at 50 µW for all measurements. The Raman intensities are normalized by the silicon Raman intensity including the resonance effect [40] for each excitation energy to correct for the efficiency of the detection system. The multiple interference effect from the substrate is accounted for [41] by using the dielectric functions and refractive indices of monolayer and bulk $WS_2$ from the literature [19,42]. Reflectance contrast spectra were obtained by measuring the reflectance spectra of the sample and the bare substrate using a supercontinuum laser (Fianium sc-400) and calculating [$R$(sample)−$R$(substrate)]/[$R$(sample)+$R$(substrate)]. The spectral resolution ranged between 0.3 cm$^{-1}$ (1.96 eV) and 1.5 cm$^{-1}$ (3.82 eV).

### 3. Results

Monolayer $WS_2$ consists of a tungsten layer sandwiched between two sulfur layers, forming a so-called tri-layer (TL) as shown in Fig. 1a. The point group corresponds to $D_{6h}$ for bulk, $D_{3h}$ for odd number of layers, and $D_{3d}$ for even number of layers [43,44]. Although the mode designations differ for even and odd number of layers, we will use those for bulk $WS_2$ for convenience. Fig. 1b shows the reflectance contrast spectra of 1 to 4 TL $WS_2$ [19,21,38]. Characteristic features corresponding to the A, B and C exciton states are observed at ~2.0 eV, ~2.4 eV and ~2.8 eV, respectively [5,19,45]. As the number of layers increases, the exciton states shift slightly due to the changes in the electronic band structure [46,47]. The separation between the A and B exciton states is rather large (~0.4 eV) compared to $MoS_2$ or $MoSe_2$ due to different size of transition metal atoms [47,48], allowing separate investigation of the resonance effects due to these two exciton states.

Fig. 1c and d show the Raman spectra as a function of the number of layers taken with the 2.54-eV excitation. See Fig. S1 and S2 for similar data for other excitation energies. The main $E_{2g}^1$ and $A_{1g}$ modes shift, and the separation between them increases with the thickness [25,29,38]. There are other smaller features corresponding to either multi-phonon scattering or those modes that are forbidden in the first order. These peaks also show slight dependence on the thickness. The relative intensities and the line shapes of these smaller peaks depend strongly on the excitation energy. Fig. 1e and f compare the Raman spectra of 4TL $WS_2$ for different excitation energies. See Fig. S3-5 for similar data for other thicknesses.

The signal due to two-phonon scattering via acoustic phonons at the $M$ point of the Brillouin zone (2LA) appears ~5 cm$^{-1}$ below the main $E_{2g}^1$ mode and becomes relatively stronger for excitation energies (1.96, 2.33, and 2.41 eV) near resonances with the A and B exciton states [29,35]. The 2LA signal shows strong polarization dependence whereas $E_{2g}^1$ is isotropic (See Fig. S6). Since the 2LA signal sometimes overwhelms the $E_{2g}^1$ peak, the spectra measured in cross polarization, in which the 2LA signal vanishes, should be used to isolate the $E_{2g}^1$ signal which is isotropic. The $E_{2g}^1$ peak shows strong enhancement near resonance with the C exciton state, whereas the $A_{1g}$ peak exhibits resonance enhancement with A, B, and C exciton states (See Fig. S7). The different resonance behaviors have been explained in terms of the correlation between the phonon vibration directions and the atomic orbitals that contribute to the exciton states [23,33,35] or so-called quantum interference between electronic transitions at different parts in the Brillouin zone. The small shoulder on the lower-frequency side of the $A_{1g}$ peak is due to Davydov splitting as reported earlier [49]. The weaker features also show enhancement depending on the excitation energy. The peak at 297 cm$^{-1}$ is close to the $E_{1g}$ mode which is forbidden in backscattering for 1TL [50]. A multi-phonon scattering mode has been ascribed to this peak by another group [29,51]. A close inspection reveals that this peak is significantly sharper in cross polarization than in parallel polarization (See Fig. S8), which implies that at least one more mode contributes to this peak. In cross polarization, a relatively sharp peak appears only for 2TL or thicker samples. We ascribe the component that survives in cross polarization to the $E_{1g}$ mode and the remaining signal to possible multi-phonon scattering. The $A_{2u}$ mode at 441 cm$^{-1}$ is Raman inactive for bulk or 1TL [50,52] but become partially Raman active for other thicknesses due to Davydov splitting [34,49,50,53]. However, this peak appears clearly for 1TL near the resonance with the C exciton state, which is similar to the case of MoS$_2$ and MoSe$_2$ [28,33,34,52]. The other weaker peaks are ascribed to multiple-phonon scattering [29,51].

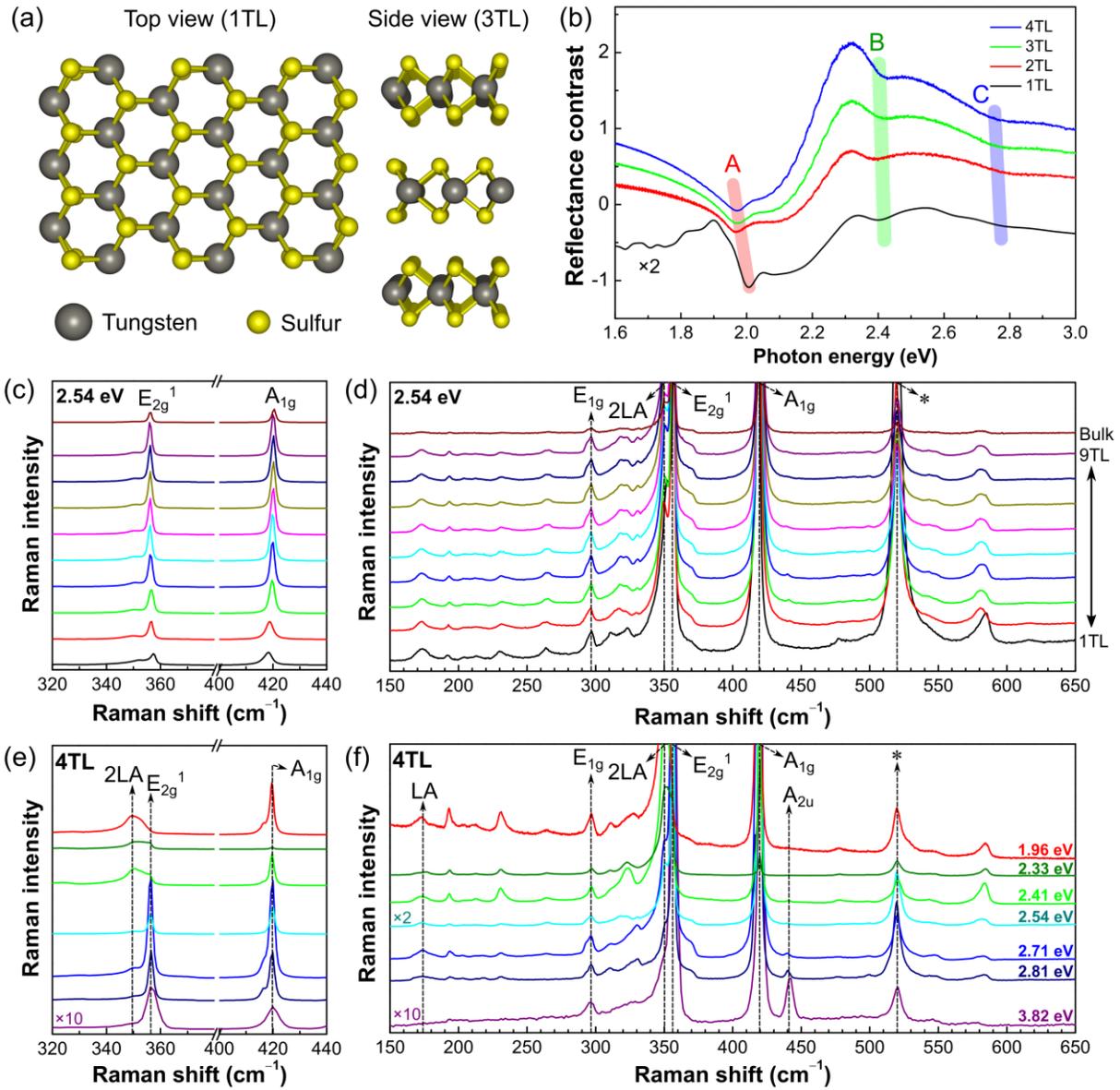

**Fig 1.** (a) Crystal structure of monolayer (top view) and 3TL (side view) WS$_2$. (b) Reflectance contrast spectra of 1TL to 4TL. Thickness dependence of Raman spectrum of (c) main peaks ($E_{2g}^1$ and $A_{1g}$) and (d) weaker peaks measured with 2.54 eV excitation. Excitation energy dependence of Raman spectrum of 4TL WS$_2$ for (e) main peaks and (f) weaker peaks. The silicon signal at 520 cm$^{-1}$ is marked by *.

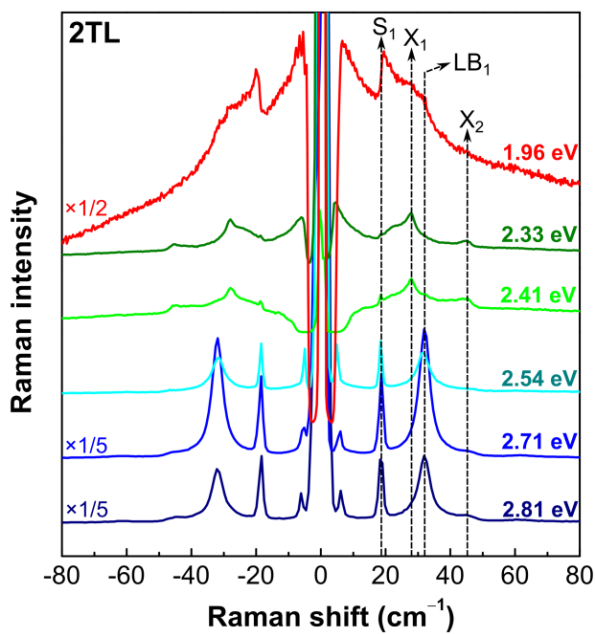

**Fig 2.** Low frequency (interlayer vibration) Raman spectra of 2TL $WS_2$ measured with six different excitation energies.

Fig. 2 shows the low-frequency Raman spectra of 2TL $WS_2$ measured with six different excitation energies. See Fig. S9 for similar data for other thicknesses. The spectral features vary greatly depending on the excitation energy, although the peak positions of the phonon modes are fixed. Some of the peaks are observed only for some specific excitation energies. The shear mode ($S_1$) at 19 cm$^{-1}$ and the layer-breathing mode ($LB_1$) at 32 cm$^{-1}$ are clearly enhanced for higher energy excitations. Two additional features, labelled $X_1$ at 28 cm$^{-1}$ and $X_2$ at 46 cm$^{-1}$, are observed for some excitation energies. Careful examination of the intensities of these two peaks reveal that they seem to be enhanced near resonances with exciton states (see Fig. S10), although they are not always resolved. These two peaks appear in all thicknesses, including 1TL, and their positions do not have dependence on the thickness (see Fig. 3a-c). The polarization dependence is different: $X_1$ disappears in cross polarization but $X_2$ does not (see Fig. S11). A similar peak, so-called X-peak, was reported for $MoS_2$ when the excitation energy was close to the A and B exciton states [28]. The origin of this peak has not yet been identified. In an effort to identify the origin of these peaks, one can consider several possibilities. Firstly, these peaks cannot be due to phonons at high-symmetry points in the Brillouin zone because no high-symmetry point phonon in 1TL $WS_2$ has energy this small except for the acoustic phonons at the zone center which has zero energy. Second, they are not related to the excitons in a simple way because the exciton energy depends on the thickness as shown in Fig. 1b, but these peaks do not seem to depend on the thickness. One might suggest that these peaks are due to some complex excitation which is a combination of low-energy acoustic phonon(s) and some low-energy quasi-particle with momentum related to the excitons. More studies are needed to elucidate this puzzling phenomenon.

In addition to the sharp peaks, there is a rather broad peak centered at the origin when the excitation energy is near the A (1.96 eV) or B (2.33 or 2.41 eV) exciton resonances. This 'central peak' was also observed for $MoS_2$ and was ascribed to acoustic phonon scattering mediated by the exciton states broadened by the inhomogeneity in the sample [28].

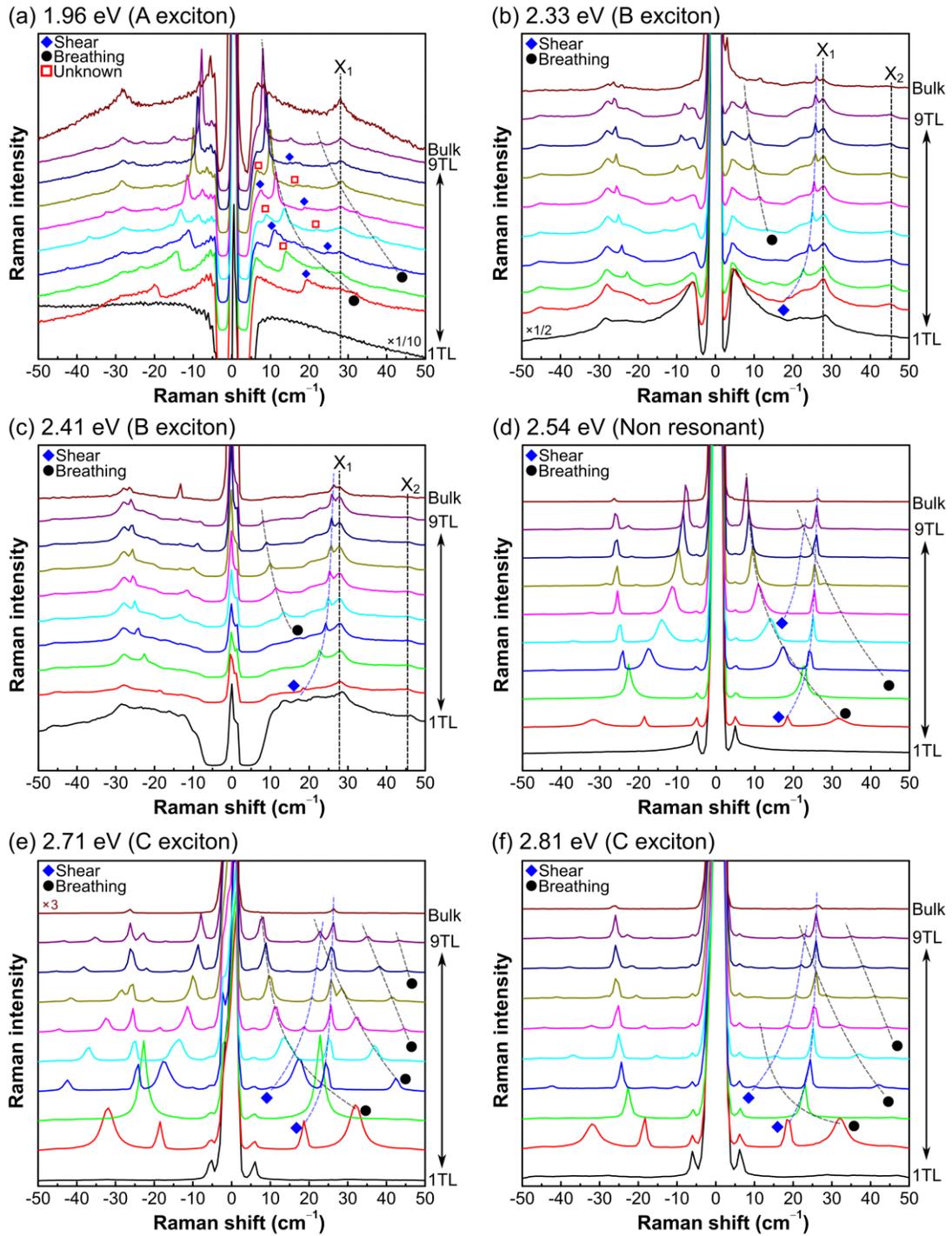

**Fig 3.** Thickness dependence of low-frequency Raman spectra for six different excitation energies. In addition to the shear and layer-breathing modes, some unknown modes are indicated. The curves are guides to the eye.

Fig. 3 shows the low-frequency Raman spectra of few-layer $WS_2$ measured with six different excitation energies. The spectral positions of the shear and layer-breathing modes depend sensitively on the number of layers. As in other layered materials, the shear modes blue shift whereas the layer-breathing modes red shift as the number of layers increases [26–28,34,39,53,54]. For 2H-type TMDs such as $MoS_2$ or $WS_2$, the point group symmetry for odd number of layers is $D_{3h}$ whereas that for even number of layers is $D_{3d}$. In backscattering geometry with the laser incident normal to the layers, (N-1)/2 doubly degenerate shear modes and (N-1)/2 layer-breathing modes are Raman active for odd number of layers. For even number of layers, the same numbers are N/2. The remaining modes are either forbidden in the backscattering geometry or infrared active (Raman inactive). Interestingly, for the 1.96 eV excitation, a number of peaks (marked by red squares in Fig. 3a) that do not correspond to the allowed shear or layer-breathing modes appear.

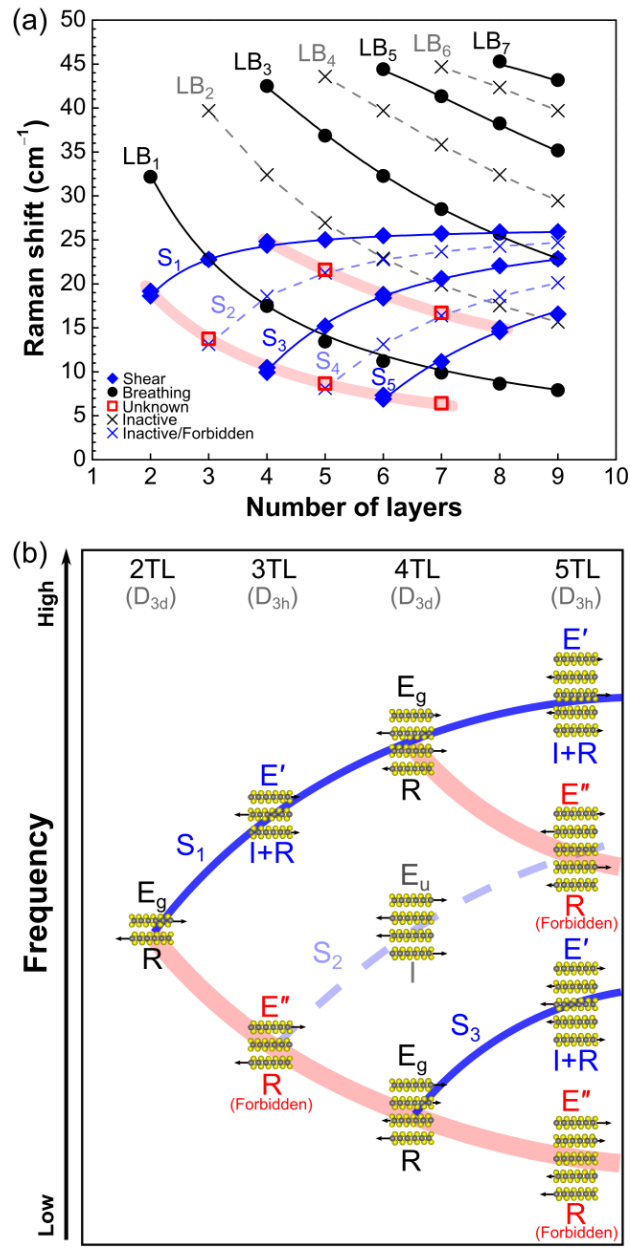

**Fig 4.** (a) Position of interlayer vibration modes as a function of the number of layers. The experimentally observed Raman active modes are shown in solid symbols, and Raman inactive or forbidden modes obtained from calculations are indicated by '×'. Red squares are experimentally observed peaks that correspond to Raman forbidden modes. (b) Schematic vibration diagram of shear modes for 2TL to 5TL $WS_2$. The mode designations for each thickness are shown. $E_g$ modes are Raman active, $E_u$ infrared active (Raman inactive), $E'$ Raman and infrared active, and $E''$ Raman active but forbidden in backscattering.

**Table 1.** Force constant of five (WSe$_2$, MoS$_2$, MoSe$_2$, MoTe$_2$ and WS$_2$) 2H-TMDs. In-plane ($K_x$) and out-of-plane ($K_z$) force constants are obtained by linear-chain-model calculations.

| Material | In-plane ($K_x$) | Out-of-plane ($K_z$) |
| --- | --- | --- |
| WS$_2$ | $2.99 \times 10^{19}$ N/m$^3$ | $9.10 \times 10^{19}$ N/m$^3$ |
| WSe$_2$ [26] | $3.07 \times 10^{19}$ N/m$^3$ | $8.63 \times 10^{19}$ N/m$^3$ |
| MoS$_2$ [26] | $2.72 \times 10^{19}$ N/m$^3$ | $8.62 \times 10^{19}$ N/m$^3$ |
| MoSe$_2$ [34] | $2.92 \times 10^{19}$ N/m$^3$ | $8.73 \times 10^{19}$ N/m$^3$ |
| MoTe$_2$ [53] | $3.44 \times 10^{19}$ N/m$^3$ | $7.83 \times 10^{19}$ N/m$^3$ |

## 4. Discussion

The positions of the low-frequency shear and layer-breathing modes are summarized in Fig. 4a. The experimentally observed shear and layer-breathing modes are fitted to the linear chain model (LCM) [26,27,39] calculations to obtained the solid curves. The in-plane and out-of-plane inter-layer force constants obtained from this fitting procedure are summarized in Table 1. The in-plane force constant ($K_x$) of WS$_2$ is similar to those for other TMDs but the out-of-plane force constant ($K_z$) is somewhat larger than the others. Using these force constants, the frequencies of Raman inactive or forbidden modes are calculated. It turns out that the 'unknown' (red squares) peaks actually correspond to the $E''$ shear modes that are Raman active but forbidden in backscattering (see Fig. 4b). The appearance of the $E''$ shear modes for the 1.96 eV excitation may be attributed to resonance-induced breaking of the local symmetry. However, polarized Raman measurements reveal that it is much more complicated than simple symmetry breaking. The shear modes with E symmetry should be isotropic on the lattice plane. For example, for the 2.54 eV (off-resonance) excitation, the intensity of the shear modes are almost identical between parallel (XX) and cross (XY) polarizations, whereas the layer-breathing modes disappear in cross polarization.

However, the lowest frequency shear modes, be it allowed ($E_g$) or forbidden ($E''$), disappear in cross polarization when the 1.96-eV excitation is used. (see Fig. S11 for polarized Raman data). This surprising polarization behavior implies that even the 'allowed' modes cannot be described as simple shear modes.

It should be noted that the lowest shear modes have pronounced asymmetric line shapes that are indicative of Fano resonance, which is characterized by the Breit-Wigner-Fano (BWF) line shape [55–57], given by [57]

$$I(\omega) = I_0 \frac{[1+2(\omega-\omega_0)/(q\Gamma)]^2}{[1+4(\omega-\omega_0)/\Gamma^2]}, \quad (1)$$

where $\Gamma$ is the broadening parameter, $1/q$ is the coupling constant, $\omega_0$ is the peak frequency of uncoupled mode, and $\Delta\omega$ (=$\Gamma/2q$) is the frequency shift due to Fano resonance [57]. Fano resonance occurs when a discrete excitation is in resonance with a continuum. In the case of graphene, the shear mode exhibits Fano resonance due to resonance with low-energy electronic excitations owing to the gapless Dirac dispersion of the electronic bands [56] whereas the main G band phonon shows Fano resonance in the undoped limit due to resonance with exciton-like excitations [55]. In the current case, there is no obvious candidate for the continuum excitation. Although the broad exciton lines may suggest that the exciton states themselves might serve as a 'continuum', recent reports that the homogeneous linewidth of the excitons in TMDs is very narrow [58–60], indicating that the commonly observed broad exciton linewidth is mainly due to inhomogeneous broadening that would not qualify as the 'continuum' for Fano resonance. On the other hand, it has been reported that excitons in transition metal dichalcogenides have a peculiar dispersion as a function of the center-of-mass momentum [61]. Therefore, excitation of an exciton to a higher energy state might be the continuum excitation which might couple with the shear mode to result in Fano resonance. Table 2 summarizes the BWF line shape parameters for 2 to 4TL WS$_2$ obtained by fitting the spectra to Eq. (1) along with other Raman peaks (see Fig. S12). The coupling coefficient $1/q$ increases with the number of layers. Since the A exciton state redshifts with the number of layers and comes closer to 1.96 eV as the number of layers increases. Therefore, one can say that Fano resonance gets stronger as the excitation energy become more resonant with the A exciton state. Fano resonance is not seen for B or C exciton resonances. The fact that the peculiar polarization behaviors and Fano resonance of the lowest shear modes appear simultaneously suggests that Fano resonance in fact modifies the symmetry of the shear modes. Furthermore, since these effects occur only for the 1.96 eV excitation, the A exciton resonance should be responsible for these unusual phenomena. One might need to consider a compound excitation comprising the shear mode, the A exciton state and/or Fano resonance. Further studies are needed to elucidate the detailed mechanism of this phenomenon.

**Table 2.** Parameters of Breit-Wigner-Fano (BWF) line shape given by Eq. (1) for the lowest-frequency shear mode peak measured with 1.96 eV excitation.

| Number of layers | $\Gamma$ | $1/q$ | $\Delta\omega$ (cm$^{-1}$) |
|---|---|---|---|
| 2TL | 3.52 | 0.68 | 1.20 |
| 3TL | 2.04 | 0.79 | 0.82 |
| 4TL | 1.81 | 0.93 | 0.84 |

## 5. Conclusions

Raman spectra of few-layer WS$_2$ were measured with up to seven different excitation energies. The main $E_{2g}^1$ and $A_{1g}$ modes shift with thickness and exhibit resonant enhancement near exciton states. Some of the weaker features are relatively enhanced for some excitation energies. The two-phonon acoustic phonon scattering (2LA) signal close to the $E_{2g}^1$ peak is stronger than the main $E_{2g}^1$ peaks for lower energy excitations but can be removed by using cross polarization.

In the low-frequency region, a series of shear and layer-breathing modes are observed. In addition, *hitherto* unidentified peaks X$_1$ and X$_2$ are observed near resonances with exciton states. The positions of these peaks do not depend of the number of layers. The polarization dependences of the two peaks are different: X$_1$ vanishes in cross polarization, but X$_2$ does not. Furthermore, Raman forbidden shear modes are observed for odd number of layers for the 1.96 eV excitation. The lowest-frequency shear mode, be it allowed or forbidden, exhibits a Breit-Wigner-Fano line shape due to Fano resonance and shows a peculiar polarization dependence for 1.96 eV excitation. The correlation between Fano resonance and the peculiar polarization dependence calls for further studies.

**Acknowledgements**

This work was supported by the National Research Foundation (NRF) grant funded by the Korean government (MSIP) (NRF-2016R1A2B3008363) and by a grant (No. 2011- 0031630) from the Center for Advanced Soft Electronics under the Global Frontier Research Program of MSIP. We thank H. C. Lee for helpful discussions.

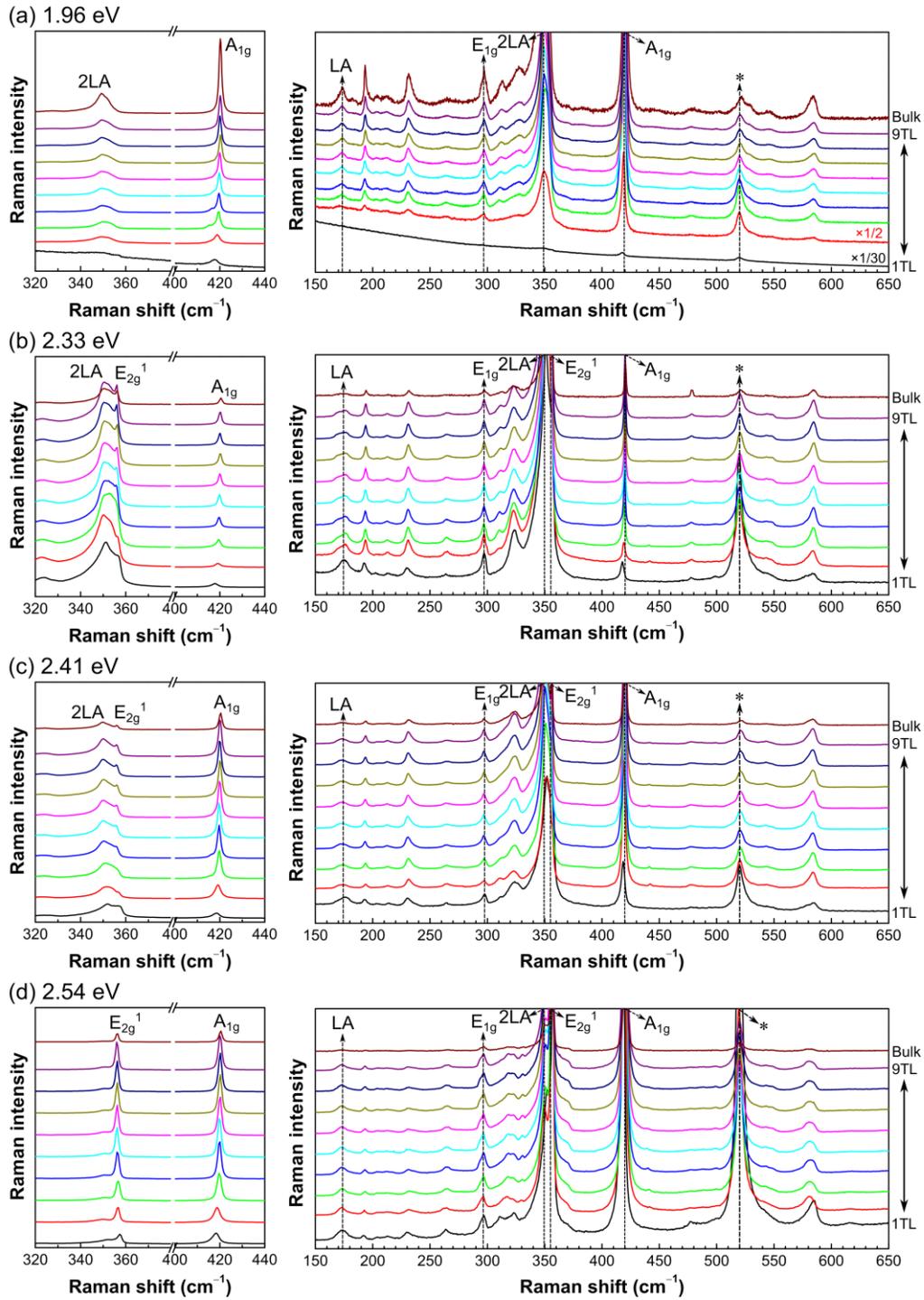

**Fig. S5**. Thickness dependence of Raman spectra for excitation energies of (a) 1.96, (b) 2.33, (c) 2.41 and (d) 2.54 eV. The mode notations are based on the case of bulk $WS_2$. Silicon peak (520 $cm^{-1}$) is indicated by '*'.

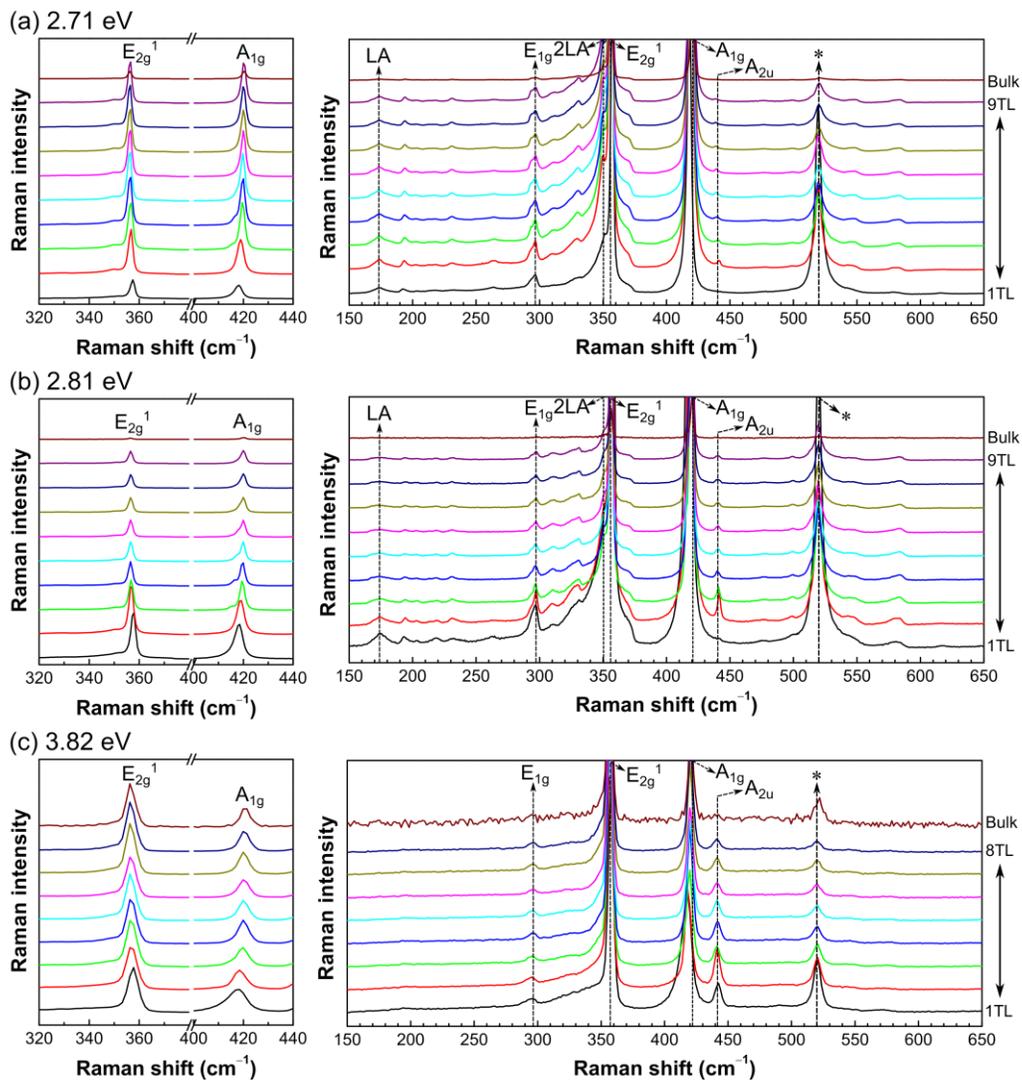

**Fig. S2**. Thickness dependence of Raman spectra for excitation energies of (a) 2.71, (b) 2.81, and (c) 3.82 eV. The mode notations are based on the case of bulk $WS_2$. Silicon peak (520 $cm^{-1}$) is indicated by '*'.

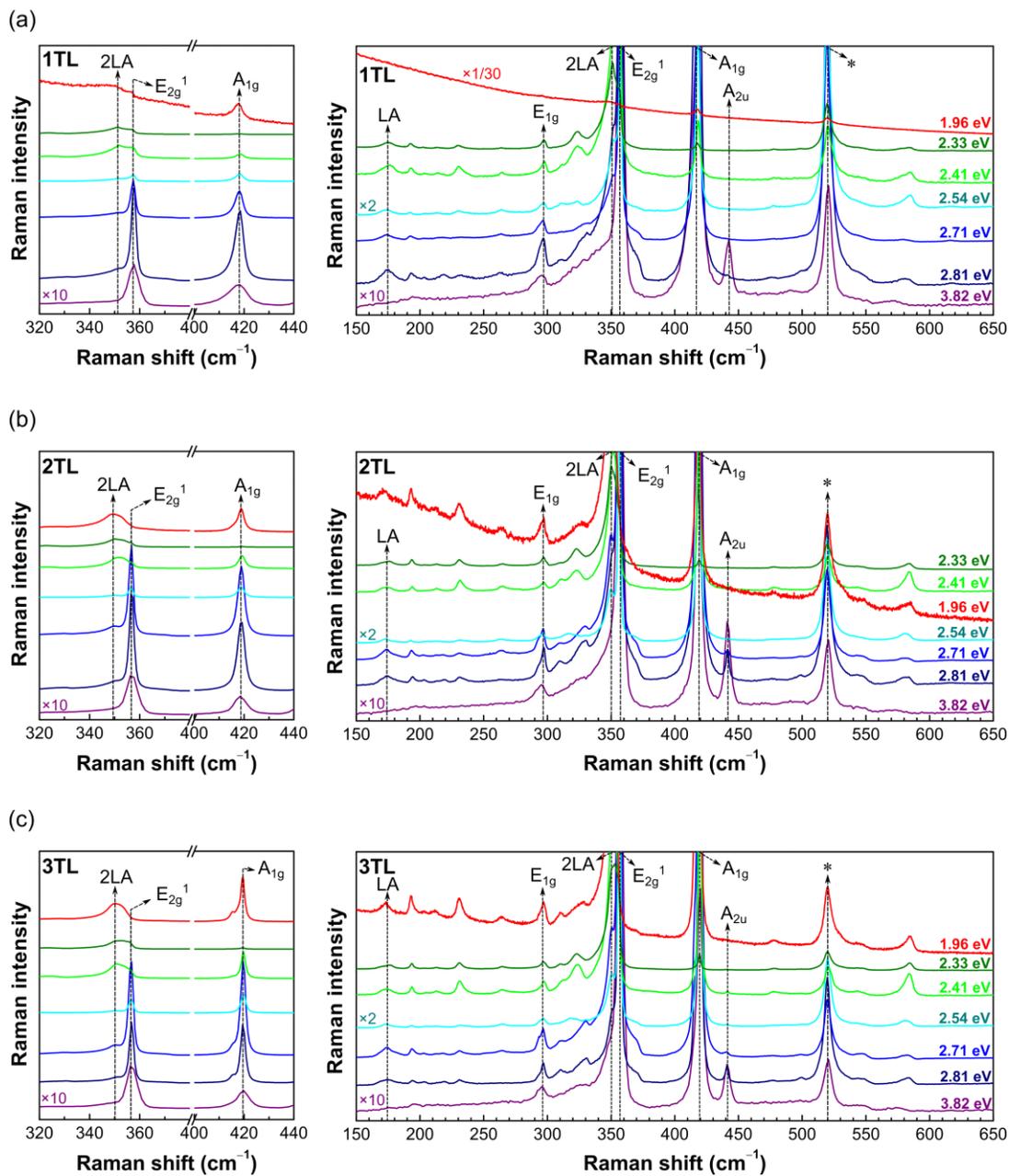

**Fig. S3.** Excitation energy dependence of Raman spectra of (a) 1TL, (b) 2TL, and (c) 3TL WS$_2$.

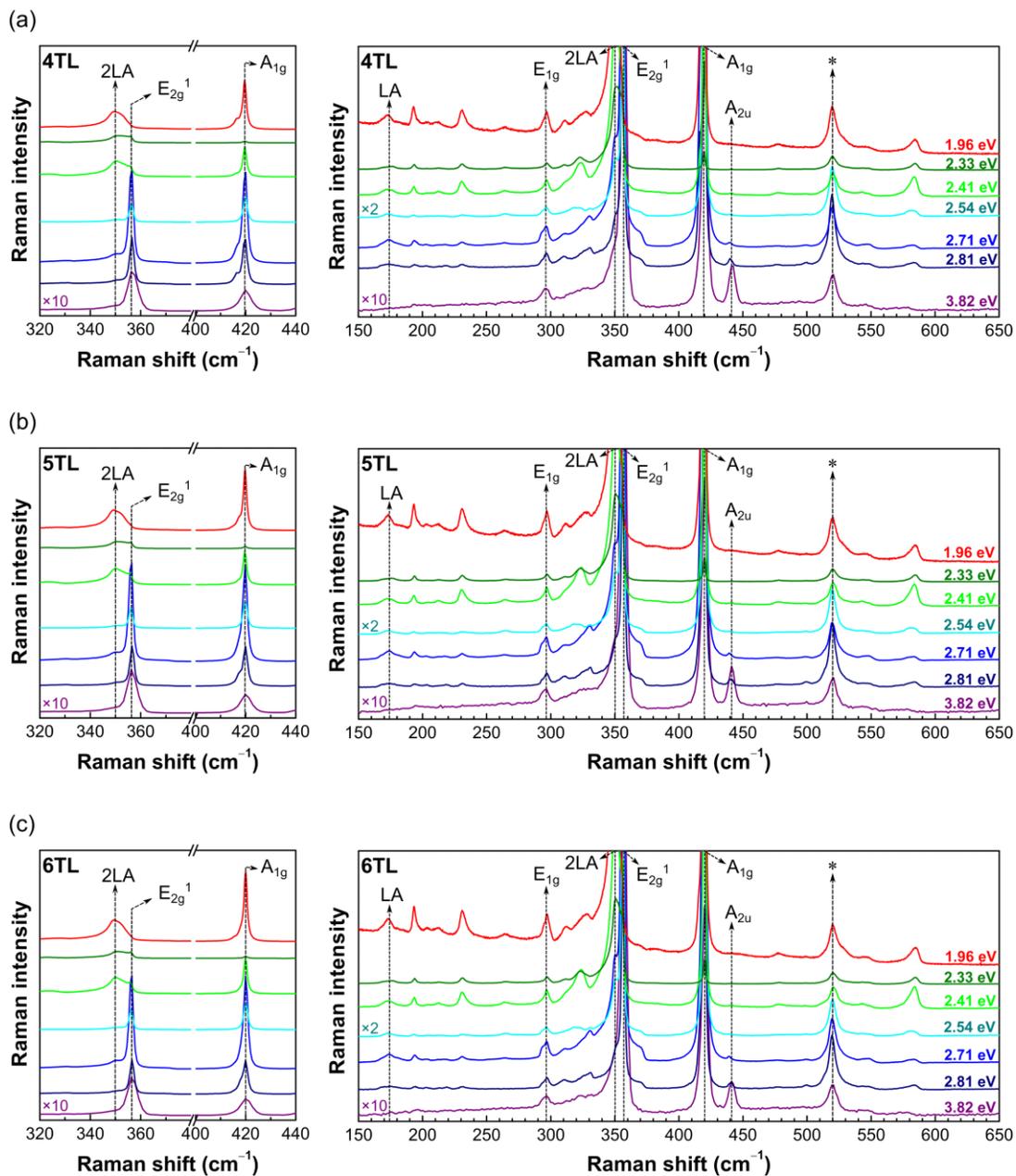

**Fig. S4**. Excitation energy dependence of Raman spectra of (a) 4TL, (b) 5TL, and (c) 6TL WS$_2$.

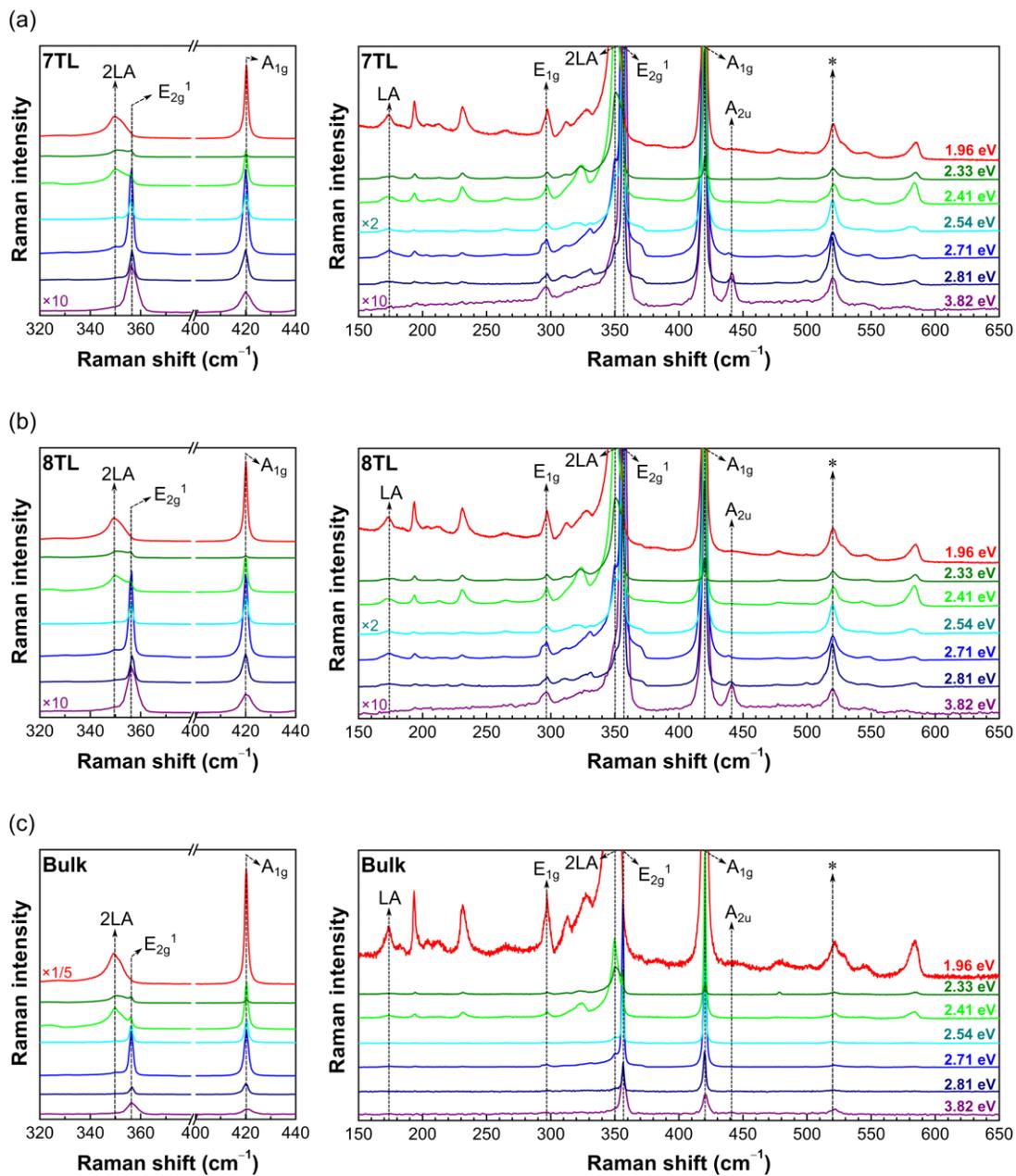

**Fig. S5**. Excitation energy dependence of Raman spectra of (a) 7TL, (b) 8TL, and (c) bulk $WS_2$.

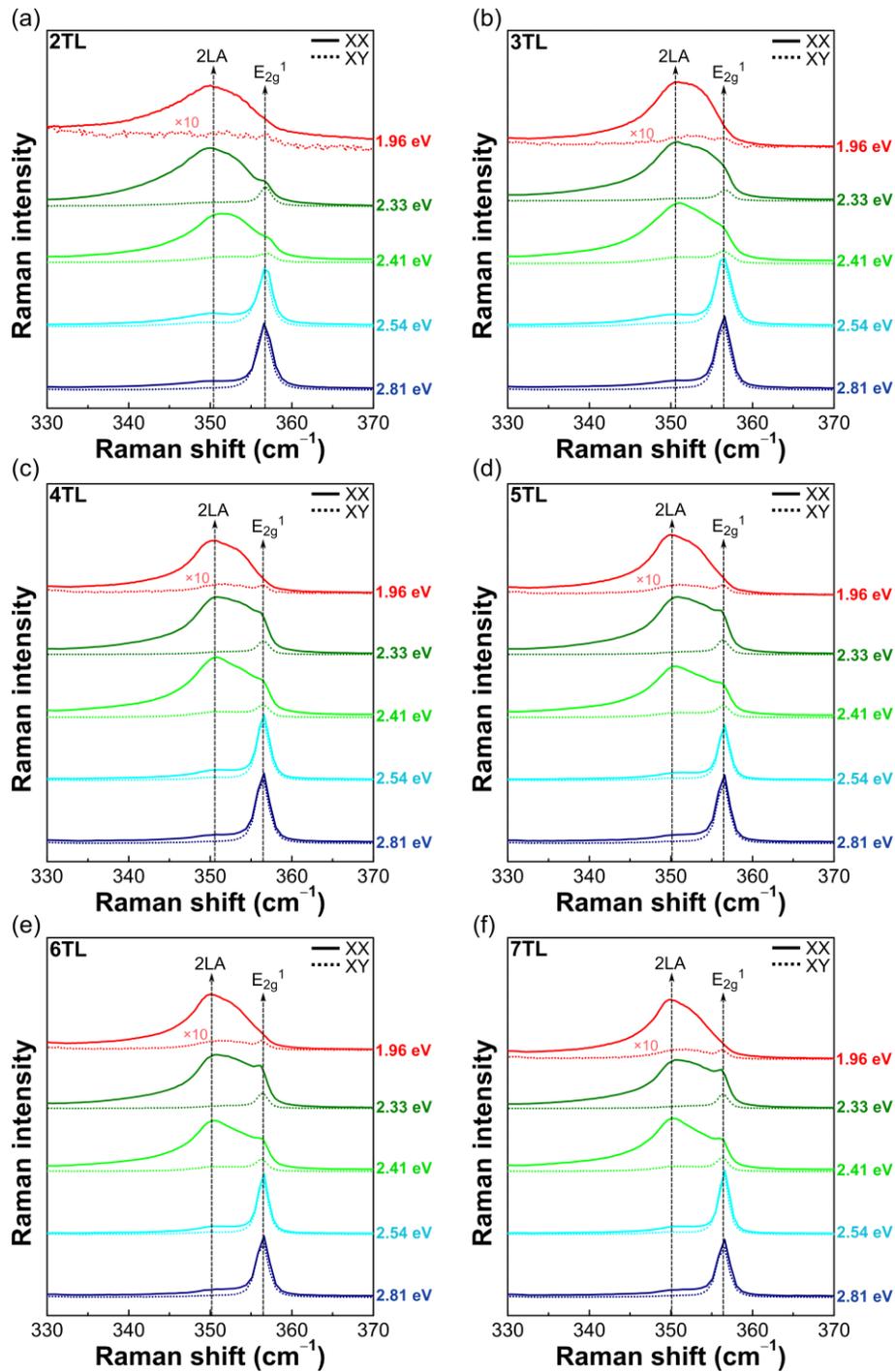

**Fig. S6**. Polarized Raman spectra of the region near the $E_{2g}^1$ mode for 2 to 7 TL WS$_2$ for five different excitation energies. Solid curves are for parallel polarization (XX), and dashed lines are for cross polarization (XY).

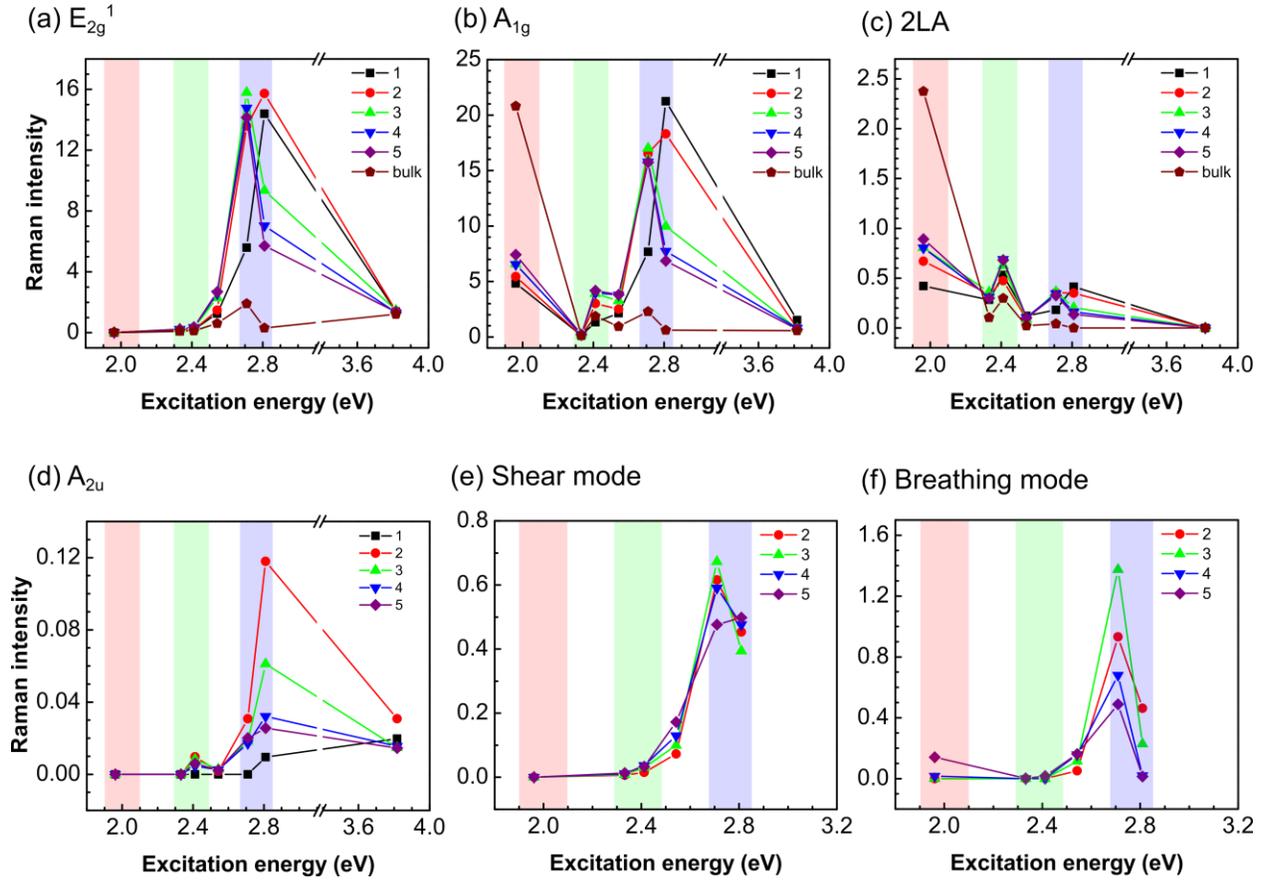

**Fig. S7**. Resonance profiles of six Raman modes for 1 to 5TL and bulk WS$_2$. The approximate energies of A, B, and C exciton states at ~2.0, ~2.4, and ~2.8 eV, respectively, are indicated.

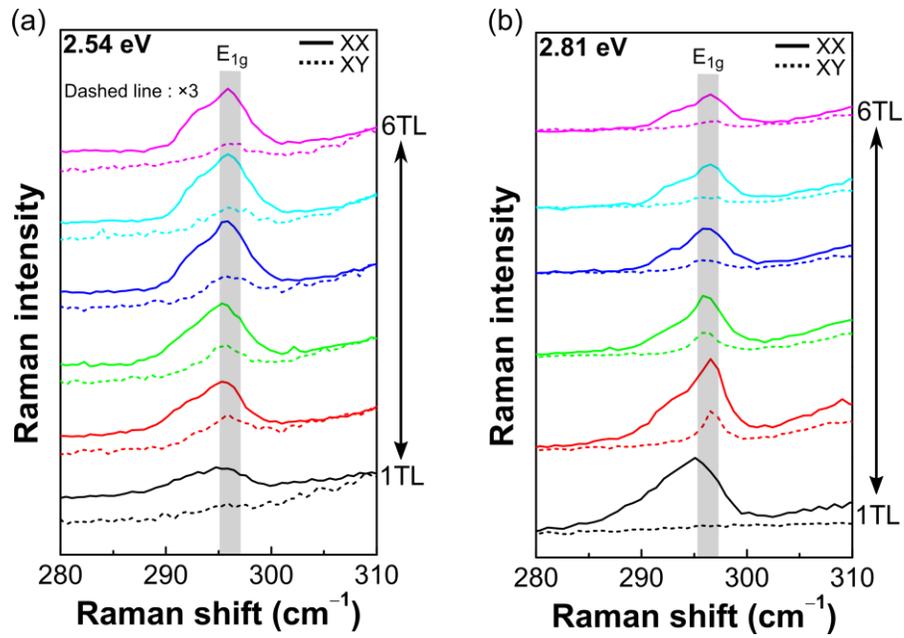

**Fig. S8**. Polarized Raman spectra of $E_{1g}$ peak measured with (a) 2.54 and (b) 2.81 eV excitation energies. The $E_{1g}$ peak is identified in cross polarization (XY) for 2TL to 6TL but does not appear for 1TL.

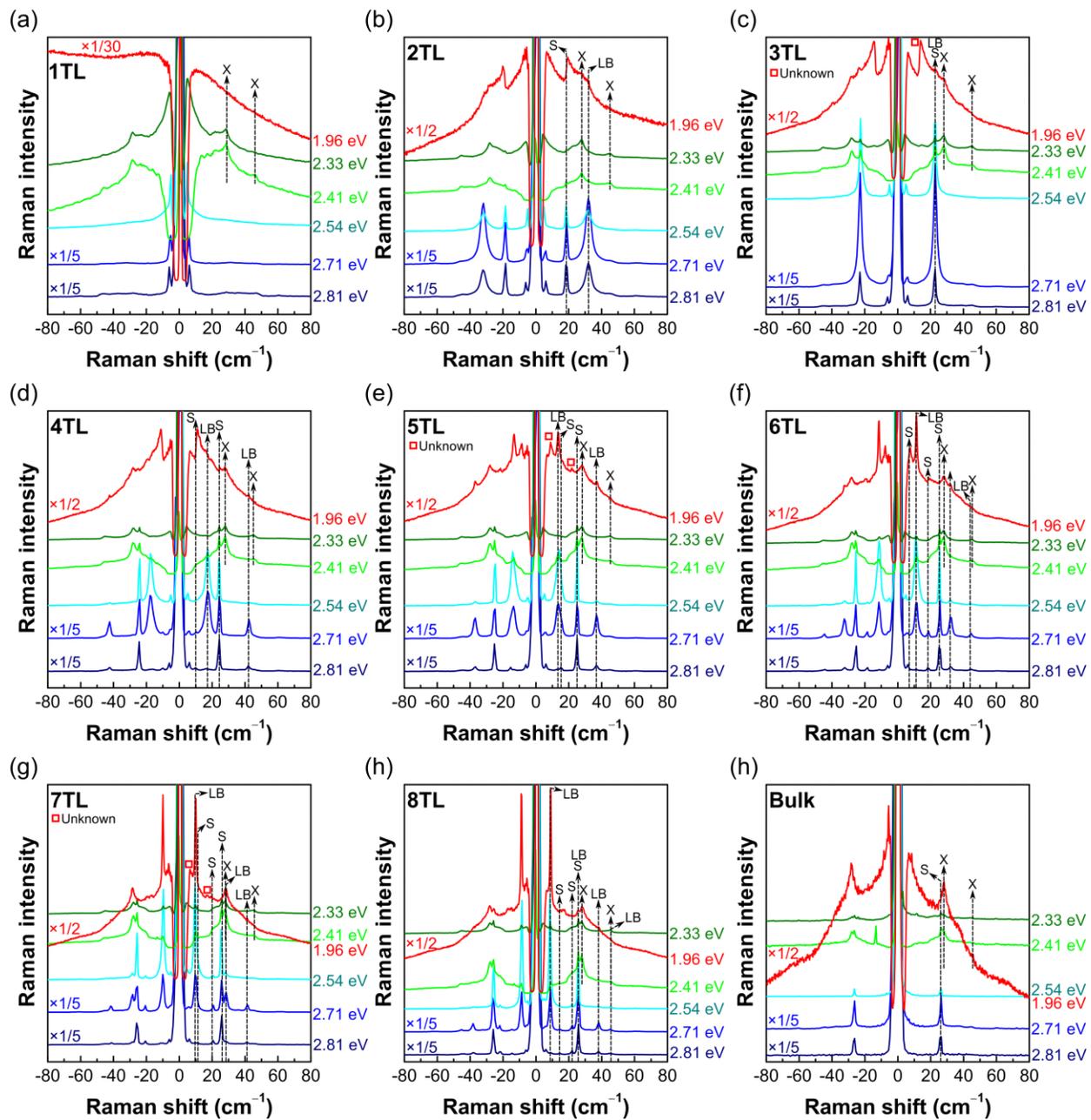

**Fig. S9**. Excitation energy dependence of Raman spectra of 1TL to 8TL and bulk $WS_2$ in the low frequency region.

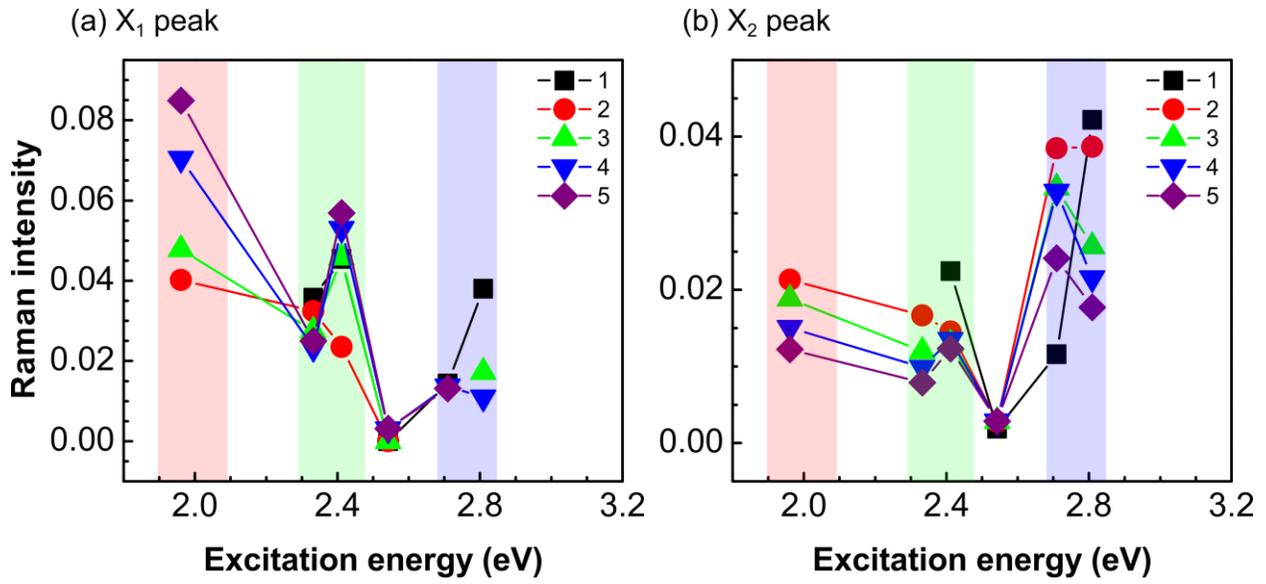

**Fig. S10**. Resonance profiles of two X peaks for 1 to 5TL and bulk $WS_2$. The approximate energies of A, B, and C exciton states at ~2.0, ~2.4, and ~2.8 eV, respectively, are indicated.

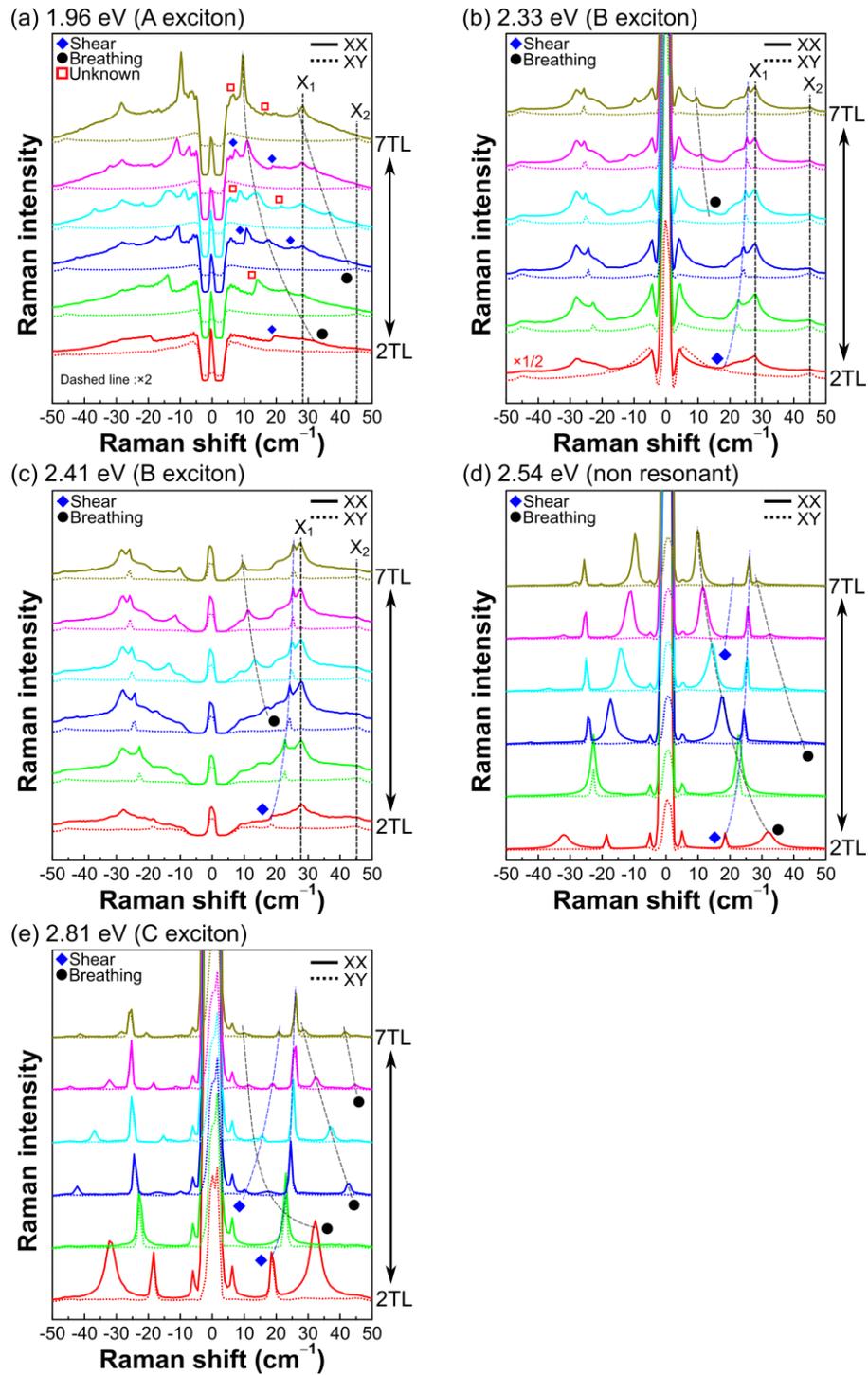

**Fig. S11**. Polarized Raman spectra in the low frequency region for 2TL to 7 TL $WS_2$ measured with five different lasers. Solid curves are for parallel polarization (XX), and dashed curves are for cross polarization (XY).

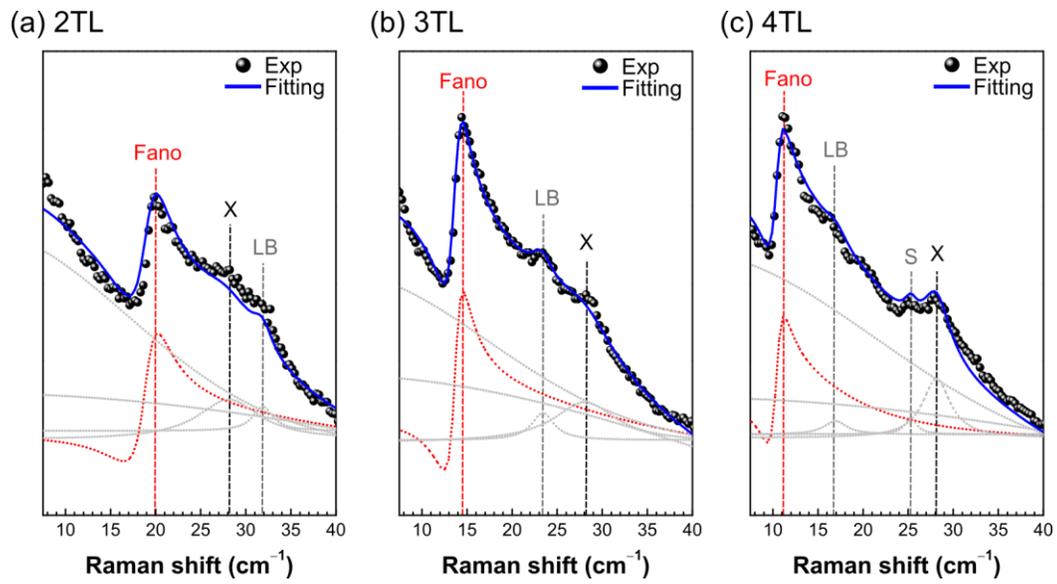

**Fig. S12**. Deconvolution of the low-frequency Raman spectra measured with 1.96 eV excitation. The lowest shear mode is fitted with the BWF line shape and the rest with Lorentzian line shapes.